\documentclass[pdflatex,sn-mathphys-num]{sn-jnl}

\usepackage{graphicx}%
\usepackage{multirow}%
\usepackage{amsmath,amssymb,amsfonts}%
\usepackage{amsthm}%
\usepackage{mathrsfs}%
\usepackage[title]{appendix}%
\usepackage{xcolor}%
\usepackage{textcomp}%
\usepackage{manyfoot}%
\usepackage{booktabs}%
\usepackage{algorithm}%
\usepackage{algorithmicx}%
\usepackage{algpseudocode}%
\usepackage{listings}%
\usepackage{lineno}
\nolinenumbers
\usepackage[symbol]{footmisc}
\usepackage[normalem]{ulem} 

\theoremstyle{thmstyleone}%
\theoremstyle{thmstyletwo}%

\theoremstyle{thmstylethree}%

\begin{document}

\title[Article Title]{
Removal of radon progeny from delicate surfaces
\footnote[3]{This work was supported by the US Department of Energy under grants DE-FG02-01ER41166 and DE-SC0012447.}}


\author*[1]{\fnm{Dmitry} \sur{Chernyak}}\email{d.n.chernyak@gmail.com}

\author[1]{\fnm{Andreas} \sur{Piepke}}

\affil*[1]{\orgdiv{Department of Physics and Astronomy}, \orgname{University of Alabama}, \orgaddress{\street{514 University Blvd.}, \city{Tuscaloosa}, \postcode{35487}, \state{AL}, \country{USA}}}

\abstract{
$\rm ^{210}Po$ $\alpha$-decay driven neutron background is a concern for many rare event search experiments. It is a difficult to control background because its radiogenic component  depends on the air exposure history of parts. In this study, we demonstrate that about half of the radon progeny $\rm ^{210}Po$ can be removed from
copper and silicon surfaces relatively easily by wiping a copper sample with acetone wetted tissue and a silicon detector with acetone soaked cotton balls.
For a copper sample we demonstrate that long-lived  $\rm ^{210}Pb$ is removed with similar effectiveness. 
For copper, allocated the longest counting time, additional wiping was found to be largely ineffective. For silicon, the removal effectiveness has large uncertainties. Additional cleaning showed a small but statistically significant effect. 
Capitalizing on this trivial cleaning step will allow experiments to relax their requirements on the allowable air
exposure time during construction, leading to cost and time savings.
}




\maketitle

\section{Introduction}\label{sec:intro}

Plate-out of radon progeny on surfaces leads to a build-up of long-lived $\rm ^{210}Pb$ with a mean live time of $\rm\tau_{Pb}=32.03\; yr$. It $\beta$-decays via $\rm ^{210}Bi$ $\rm (\tau_{Bi}=7.231\; d)$ into 
$\alpha$-unstable $\rm ^{210}Po$ $\rm (\tau_{Po}=199.637\; d)$. 
When $\rm ^{210}Po$ is present on material surfaces with a high enough interaction cross section, nuclear $\rm (\alpha, n)$ reactions can lead to unwanted neutron production. 
In many rare event search experiments, such as nEXO~\cite{nEXO:2022nam}, LEGEND~\cite{legend_2021}, JUNO~\cite{juno_2016} and XLZD~\cite{xlzd_2022}, this is an important and difficult to control  source of background events.
Neutrons are penetrating, capable of traveling deep into even large detectors. In dark matter searches their interaction with the detector medium can create signal-like events. Neutron capture on detector materials can lead to the creation of energetic $\gamma$-backgrounds. In the KamLAND experiment, the desorption of radon progeny $\rm ^{210}Bi$ from the surface of its liquid scintillator containment ballon was found to be an important background source~\cite{keefer_2015}.

Because future rare event searches have ever more challenging background requirements, control of this air-exposure history dependent background will become important for achieving the physics goals. 

The properties of the attachment of the radon progeny to surfaces and its dependence on environmental conditions have been studied in detail~\cite{chernyak_2023}. 
Coupled with neutron transport and detector acceptance calculations, the attachment properties can be translated into maximally allowable exposure durations for detector parts in a given radon-containing environment. An example is given in reference~\cite{veeraraghavan_2023} for nEXO.

Clearly, even partial removal of these background-creating radon progeny from surfaces would enable longer exposure times and with it relax time pressure on the assembly process. 
Experiments often use relatively expensive radon removal devices in their assembly laboratories to extend the allowable exposure time. The removal of  half of the surface-attached $\rm ^{210}Pb$ and $\rm ^{210}Po$ by simple wiping, as reported here,  offers a cost-effective alternative to radon removal for extending the assembly duration.
There is literature available on such removal, but in most cases,  describing harsh methods like etching and electro polishing~\cite{zuzel_2012,zuzel_2015,zuzel_2018,bunker_2020}. 
Such treatments are likely not possible for delicate components or those with technical functionality. Reference \cite{bruenner_2021} discusses different cleaning approaches for PTFE, including wiping with ethanol-soaked tissue.

The study presented here quantifies the removal fractions obtained by wiping surfaces with acetone,  a minimally intrusive approach,  which is likely allowable for most components. It focuses on copper and silicon.

\section{Measurement approach}
\label{sec:data}

As shown in~\cite{chernyak_2023}, short-lived $\rm ^{218}Po$, $\rm ^{214}Pb$ and $\rm ^{214}Bi$ are plating out on surfaces exposed to airborne radon progeny. Due to its long life, their decay product $\rm ^{210}Pb$ accumulates on surfaces. Its decay sequence is shown in~\ref{eq:decay_chain}.
\begin{equation}
^{210}{\rm Pb}\; \substack{32.0\; y\\ \xrightarrow{\makebox[0.6cm]{}}\\ \beta} \; ^{210}{\rm Bi}\;
\substack{7.23\; d\\ \xrightarrow{\makebox[0.6cm]{}}\\ \beta} \; ^{210}{\rm Po}\;
\substack{200\; d\\ \xrightarrow{\makebox[0.6cm]{}}\\ \alpha} \; ^{206}{\rm Pb}\; \label{eq:decay_chain}
\end{equation}

\begin{figure}[bt]
\includegraphics[width=8.3cm]{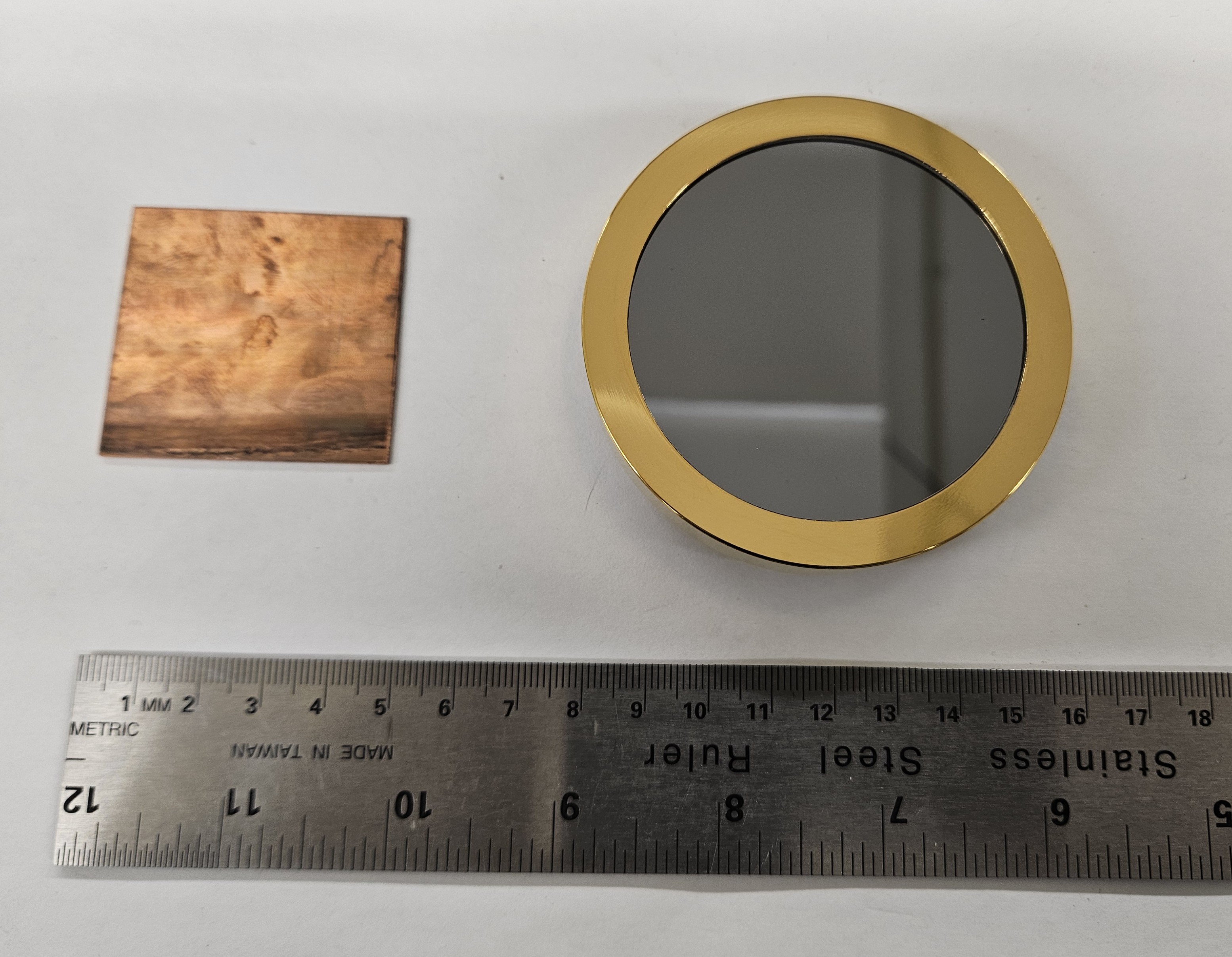}
\caption{\label{fig:samples} 
Photo of copper sample 1 and the Si detector used to study surface activities.
}
\end{figure}

To test radon progeny removal, we used two thin sheets of high-purity electrolytic EXO-200 copper~\cite{leonard_2008} that had been exposed to laboratory air from 2016 to 2024. The copper samples have a smooth, rolled finish.
We also used a Si detector, with a mirror-finish surface, with $\rm ^{210}Po$ deposited on its surface. It had migrated there during repeated Po measurements of various high-activity samples.
A different Si detector of the same size and type was used in the study of the copper sample. Two of the samples are shown in figure~\ref{fig:samples}.

Due to the long exposure of the two copper samples (about 14 $\rm ^{210}Po$ mean live times), secular equilibrium within decay chain~\ref{eq:decay_chain} is ensured. 
The effective exposure duration of the Si detector is not known.
For this study, exact knowledge of the duration and circumstances of the exposure is not needed. 
Its goal is to evaluate the efficiency of $\rm ^{210}Pb/\rm ^{210}Po$ removal by wiping with acetone-wetted Kimwipes, a method that can be applied even to delicate surfaces. 
The choice of solvent was based on the recommendation of the silicon detector manufacturer ORTEC. Their detectors can withstand such treatment and still remain functional.
This test was performed by comparing ``before'' and ``after'' $\alpha$-spectroscopic measurements of the $\rm ^{210}Po$ surface activities of the different samples. 
Due to the short range of $\alpha$-particles in matter, this approach ensures that only surface activities contribute.

Focusing on the $\rm ^{210}Po$ decay alone still allows to determine the removal of its long-lived progenitor $\rm ^{210}Pb$ by means of the time dependence of the event rate, at least for the copper samples.
If $\rm ^{210}Pb$ is left behind, $\rm ^{210}Po$ will grow back, effectively counteracting the cleaning.
Due to its short live time, $\rm ^{210}Bi$ is not of interest. It will grow back into equilibrium with $\rm ^{210}Pb$ within a few weeks.

The time dependence of sequential decays, such as in~\ref{eq:decay_chain}, is governed by coupled equations. Let the number of $\rm ^{210}Pb$, $\rm ^{210}Bi$ and $\rm ^{210}Po$ atoms, present at time $\rm t$ on the surface of the sample be $\rm N_{Pb}(t)$, $\rm N_{Bi}(t)$, and $\rm N_{Po}(t)$.
%
\begin{eqnarray}
\frac{dN_{Pb}(t)}{dt} & = & - \frac{N_{Pb}(t)}{\tau_{Pb}} \label{decay_210pb}
\nonumber
\\
\frac{dN_{Bi}(t)}{dt} & = & \frac{N_{Pb}(t)}{\tau_{Pb}} - \frac{N_{Bi}(t)}{\tau_{Bi}} 
\label{decay_210bi}
\nonumber
\\
\frac{dN_{Po}(t)}{dt} & = & \frac{N_{Bi}(t)}{\tau_{Bi}} - \frac{N_{Po}(t)}{\tau_{Po}} 
\label{decay_210po}
%
\end{eqnarray}
%
We assume the following to hold for the copper samples: before cleaning $\rm ^{210}Pb$, $\rm ^{210}Bi$ and $\rm ^{210}Po$ were in secular equilibrium; had equal initial decay rates ($\rm A_{x, i}=N_{x, i}/\tau_x$). The long exposure of the copper samples justifies this approximation.
The cleaning represents a discontinuity, taken to define $\rm t=0$. 
The initial post-cleaning activities $\rm A_{Pb}(0)$, $\rm A_{Bi}(0)$ and $\rm A_{Po}(0)$ serve as boundary conditions.
We describe the time evolution of the fraction of the $\rm ^{210}Po$ activity that survives cleaning by dividing the solution of equation~\ref{decay_210po} by $\rm A_{Po, i}$:

\begin{eqnarray}
\frac{R_{Po}(t)}{R_{Po,i }} = \frac{A_{Po}(t)}{A_{Po,i }} 
      & = & \frac{A_{Po}(0)}{A_{Po, i}}  \cdot e^{-t/\tau_{Po}} + \nonumber \\ 
      &   & \frac{A_{Bi}(0)}{A_{Po, i}}  \cdot  \frac{\tau_{Bi}}{\tau_{Bi}-\tau_{Po}}\cdot \left( e^{-t/\tau_{Bi}} - e^{-t/\tau_{Po}}  \right) +  \nonumber \\
                                   &    &  \nonumber \\
                                   &    &  \frac{A_{Pb}(0)}{A_{Po, i}}\cdot  \frac{\tau_{Pb}}{\tau_{Pb}-\tau_{Bi}}\cdot \nonumber \\
                                   &    &  \nonumber \\
                                    &   &
                                   \left[ \frac{\tau_{Bi}}{\tau_{Bi}-\tau_{Po}}\cdot \left( e^{-t/\tau_{Po}} - e^{-t/\tau_{Bi}}  \right) \right. -\nonumber \\ 
                                   &  &  \nonumber \\
                                   &  & \left. \frac{\tau_{Pb}}{\tau_{Pb}-\tau_{Po}}\cdot \left( e^{-t/\tau_{Po}} - e^{-t/\tau_{Pb}}  \right)   \right]  \label{eq:po_decay} 
\end{eqnarray}
The $\rm ^{210}Po$ activity ratio equals the ratio of observed counting rates: $\rm R_{Po}/R_{Po,i}= A_{Po}/A_{Po,i}$, because the same sample was counted with the same detector in the same geometry.
Due to the assumption of secular equilibrium before cleaning, $\rm A_{Pb,i}=A_{Bi,i}=A_{Po,i}$, the fit of the $\rm ^{210}Po$ time dependence to equation~\ref{eq:po_decay} also determines the $\rm ^{210}Pb$ and $\rm ^{210}Bi$ activity ratios.


\begin{figure}[bt]
\includegraphics[width=13cm]{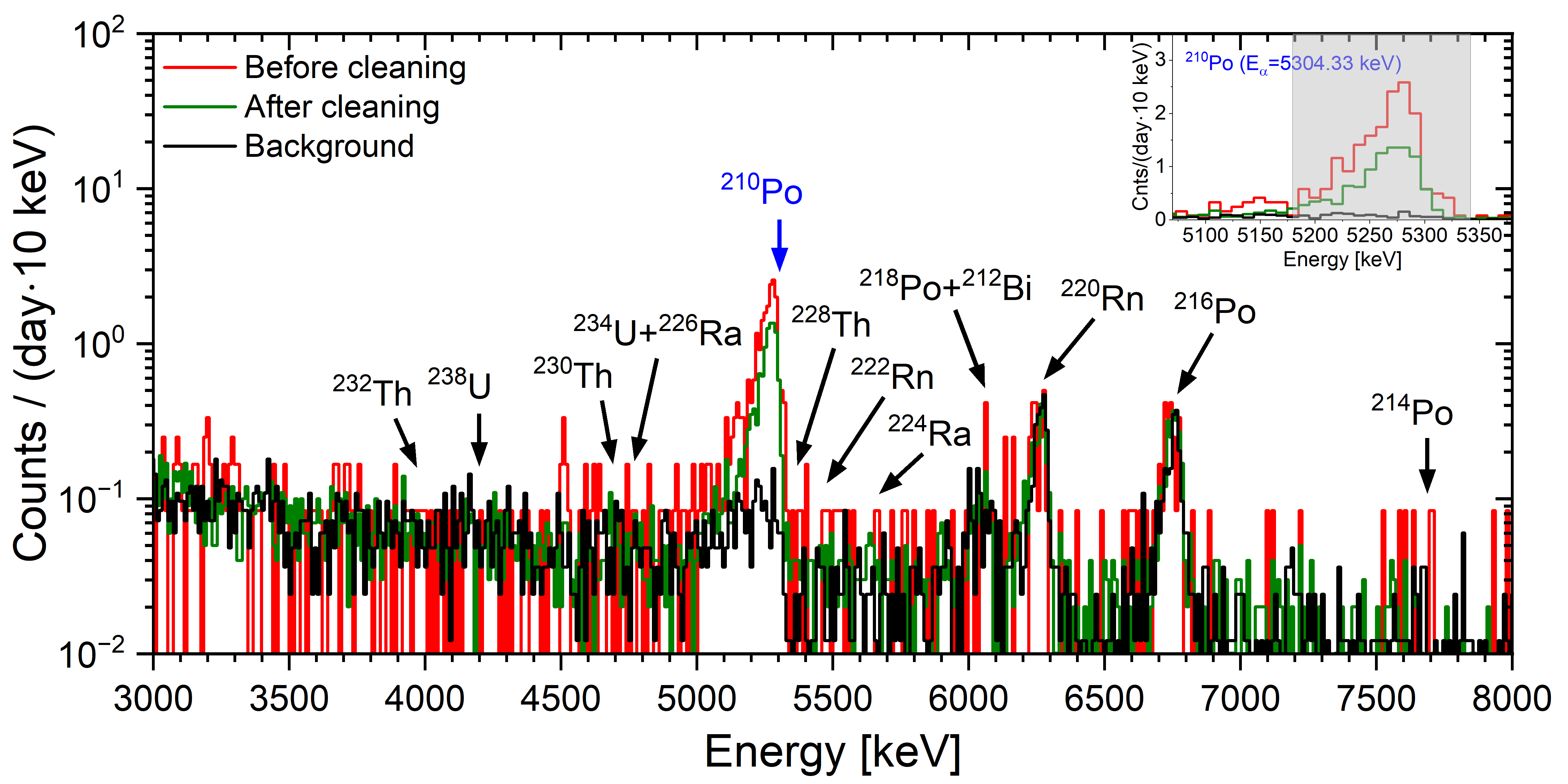}
\caption{\label{fig:po_peak} Energy spectra obtained with uncleaned  (red, 12 days), cleaned (green, 100 days) copper sample 1 and without the sample (black, 83 days). The inset shows the energy range of the $\rm ^{210}Po$ peak. A peak is visible at the expected energy. The shaded area corresponds to the chosen integration range.
}
\end{figure}
The effectiveness of $\rm ^{210}Po$ removal, by wiping with acetone, was studied using two rectangular copper and one cylindrical silicon sample. Their respective dimensions are given in table~\ref{tab:rate_results}. Two of the three samples are shown in figure~\ref{fig:samples}.
Making use of the long ``natural'' exposure of the copper sheets, the $\rm ^{210}Po$ surface activity of copper sample 1 was repeatedly measured over an extended time period, allowing to quantify the removal of $\rm ^{210}Pb$ too. 

$\rm ^{210}Po$-surface radioactivity on the samples was detected using an ORTEC low background ULTRA ENS-U3000 Si detector with $30\; \rm cm^2$ active area, operated in a vacuum chamber. 
During data collection, the chamber was continuously pumped to avoid energy loss of the $\alpha$-particles.
An oil-free Pfeiffer HiScroll 12 scroll pump was used for this purpose. 
An ORTEC Alpha Mega integrated measurement system  was used to collect the data. The energy scale of the device was calibrated with an Eckert $\&$ Ziegler $\alpha$-source, containing $^{239}$Pu, $^{241}$Am, and $^{244}$Cm radioactivity.
In case of copper, the samples were placed about 12 mm from the silicon detector. 
The counting efficiency for this arrangement was estimated with a GEANT4 simulation to be $\rm \varepsilon_{Po}=0.27$. 
The Monte Carlo code was tuned to reproduce data taken with a calibrated $\alpha$-source.
Details of the Monte Carlo simulation and its verification can be found in reference~\cite{chernyak_2023}, using the same detector.
However, the detection efficiency does not enter the analysis of the time dependence of the $\rm ^{210}Po$ event rate. 

\begin{table*}[h!bt]
    \centering
    \begin{tabular}{|c|c|c|c|c|c|c|} \hline
         & \multicolumn{2}{c|}{Copper 1} 
         & \multicolumn{2}{c|}{Copper 2} 
         & \multicolumn{2}{c|}{Silicon}  \\
         & \multicolumn{2}{c|}{($\rm 4.7\; cm\; \times 4.3\; cm$)}
         & \multicolumn{2}{c|}{($\rm 4.9\; cm\; \times 4.5\; cm$)}
         & \multicolumn{2}{c|}{(diameter $\rm 6.2\; cm$)} \\
         & Rate [cpd] 
         & $\Delta T$ [d] 
         & Rate [cpd] 
         & $\Delta T$ [d] 
         & Rate [cpd] 
         & $\Delta T$ [d]  \\ \hline
    Before cleaning & 
    15.8$\pm$1.2 
    & 12 
    & 7.63$\pm$0.54 
    & 31 
    & 1005$\pm$31
    & 16\\ \hline
    $\rm1^{st}$ cleaning 
    & $8.64\pm0.92$ 
    & 12
    & 4.14$\pm$0.61 
    & 15
    & 623$\pm$28
    & 2\\ 
    Removed fraction 
    & (-45.4$\pm7.1)\%$ 
    & 
    & (-45.7$\pm$8.9)\% 
    & 
    & (-38.0$\pm$3.3)\%
    & \\ \hline
    $\rm2^{nd}$ cleaning 
    & 8.41$\pm$0.35 
    & 88
    & 4.81$\pm$0.72 
    & 12
    & 519$\pm$18
    & 3\\ 
    Removed fraction 
    & (-3$\pm$11)\% 
    & 
    & (16$\pm$24)\%  
    & 
    & (-16.8$\pm$4.7)\%
    & \\ \hline
    $\rm3^{rd}$ cleaning 
    &  
    & 
    & 
    & 
    & 468$\pm$14
    & 4\\ 
    Removed fraction 
    &  
    & 
    &  
    & 
    & (-9.8$\pm$4.1)\%
    & \\ \hline
    \end{tabular}
    \caption{ $\rm ^{210}Po$ counting rates before/after cleaning for copper and silicon samples. $\Delta T$ denotes the counting time. Copper pieces were wiped with acetone-wetted Kimwipes. The Si detector was cleaned by carefully wiping it with acetone wetted cotton balls to avoid scratching. For the copper samples background was subtracted. The removal fractions are relative to the previous step.}
    \label{tab:rate_results}
\end{table*}

\begin{figure}[bt]
\includegraphics[width=13cm]{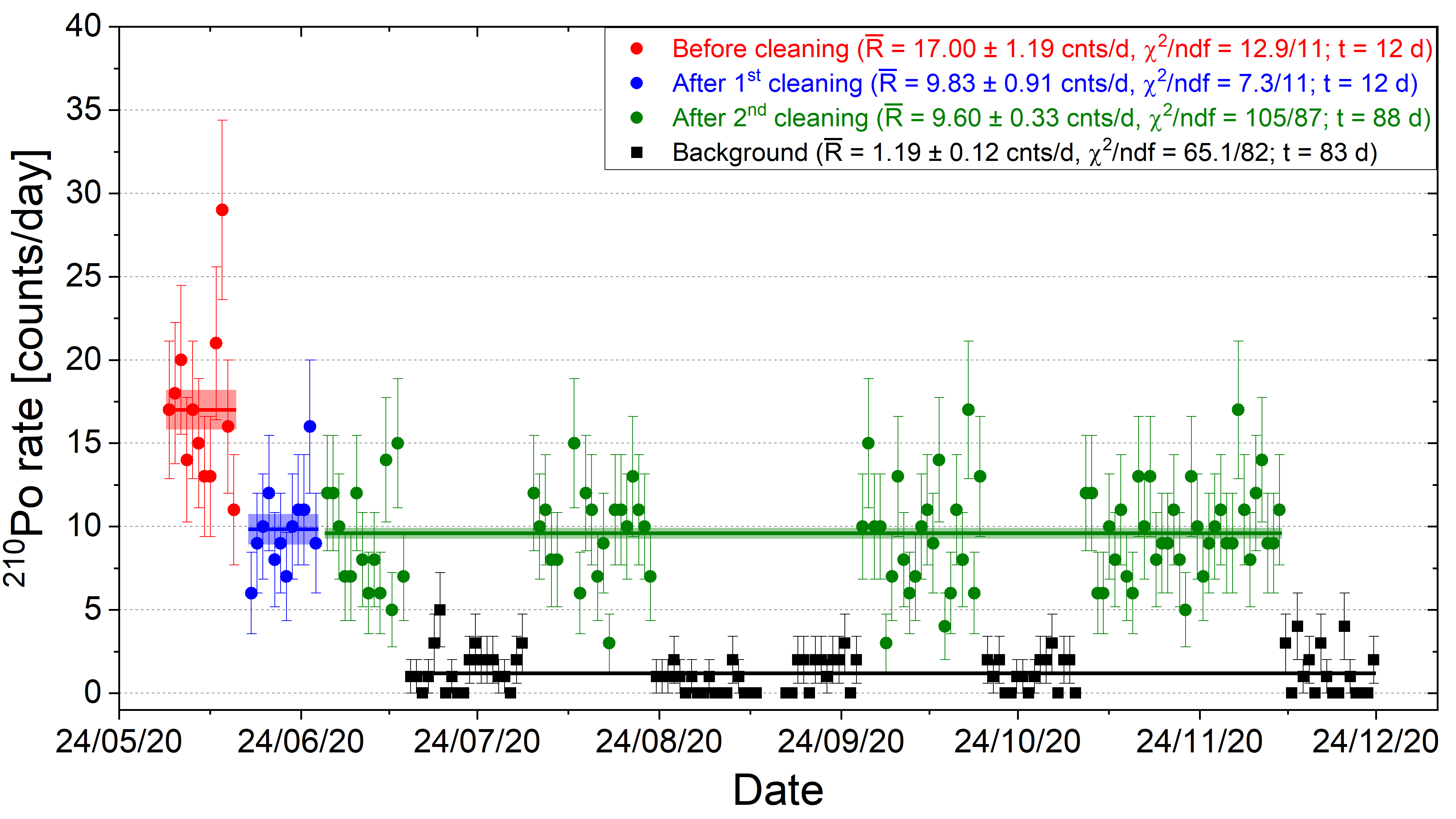}
\caption{\label{fig:time_dependence} Time dependence of the $\rm ^{210}Po$ $\alpha$-peak counting rate obtained with copper sample 1. Each point corresponds to the rate observed during one day of counting, errors are statistical. The red points were obtained with the uncleaned sample, the blue points after the initial cleaning and the green points after repeated cleaning. The black points show rates observed during background runs. The widths of the hatched bands indicate the statistical standard errors of the averages.
The x-axis gives the date in format year, month, day.}
\end{figure}

To provide a pre-cleaning normalization for the copper samples, fine particles were removed from their surfaces by blowing with boil-off nitrogen. 
For all samples, the $\alpha$-counting was broken into 1 day long intervals to allow a time-differential analysis. 

Figure~\ref{fig:po_peak} shows the time-integrated $\alpha$-spectra obtained for copper sample 1. The inset depicts the energy range around the $\rm ^{210}Po$ $\alpha$-peak. 
A peak is visible at the energy expected for $\rm ^{210}Po$. The absence of $\rm ^{224}Ra$, $\rm ^{226}Ra$, $\rm ^{232}Th$ and $\rm ^{238}U$ $\alpha$-peaks excludes surface dust as the source of activity. The other samples showed similar results.
Due to the low number of events, the time-dependent $\rm ^{210}Po$ peak integrals were determined by means of numerical integration and not fitting. 

The summed $\alpha$-spectra, shown in figure~\ref{fig:po_peak}, on the other hand, have sufficient statistics to allow peak fitting. 
In the summed spectra, the $\rm ^{210}Po$ peaks were fitted with a convolution of an exponential low-energy tail with a Gaussian distribution, taken from reference~\cite{pomme_2015}. To account for the limited statistics, a Maximum Likelihood fit was used. 
Using the fit functions resulting from the counting of the pre- and post-cleaning data, we define the ratio of function integrals from the centroid to the upper analysis edge divided by that from the lower edge to the centroid as our figure of merit for the peak asymmetry.
We obtain asymmetries of $0.356\pm 0.047$ and $0.351\pm 0.021$ for the $\rm ^{210}Po$ $\alpha$-peaks of the pre- and post-cleaning data set, respectively. There was no significant change in the low-energy tail due to cleaning. 
We conclude that the cleaning did not leave an absorbent layer.\\

\noindent
\underline{Cleaning:}\\
Following the initial counting, the copper pieces were wiped 5 times with acetone-wetted Kimwipes and counted again.
In case of the Si detector, following manufacturer instructions, the surface was cleaned by carefully wiping it multiple times with acetone wetted cotton balls to avoid scratching.

For all samples, this initial cleaning was followed by counting and a second acetone wipe.
Copper sample 1 was studied more carefully than the others to also learn about $\rm ^{210}Pb$ removal. Copper sample 1 was counted for 112 days, copper sample 2 for 58 days, and the silicon detector for 25 days. 
In case of copper sample 1, sample counting periods were alternated with background measurements to demonstrate the stability of the system. During background counting periods, the sample was kept in a sealed vacuum chamber to suppress re-attachment of radon progeny during that time.\\

\noindent
\underline{Results:}\\
$\rm ^{210}Po$ peak counting rates observed for the different samples, before and after cleaning are given in table~\ref{tab:rate_results}. In case of the copper samples the detector background rate of $1.19\pm 0.12$ cpd has been subtracted. In case of the silicon, it was the detector itself that was being observed, no detector background has to be subtracted.

Table~\ref{tab:rate_results} further gives the removal fraction, defined as the rate after cleaning minus that observed before cleaning, divided by the rate before cleaning. The removal fractions are relative to the previous step. This quantity follows the definition in reference~\cite{bruenner_2021} to allow for a direct comparison.

For all materials, close to 50\% $\rm ^{210}Po$ removal has been observed after the first cleaning. 
The reasons for the differences in  removal efficiencies between copper and silicon and how they might be related to the details of the cleaning procedures (Kimwipes vs. cotton balls) have not been investigated.
The effectiveness of the first cleaning step for the two copper samples is remarkably similar. 
For both copper samples, no statistically significant removal of $\rm ^{210}Po$ was observed for a second cleaning. 
The results agree, at least qualitatively, with those reported in reference~\cite{bruenner_2021}, table 2. The authors report that wiping PTFE discs with ethanol wetted clean room wipes resulted in the removal of close to 50\% of the $\rm ^{210}Po$ from the sample surfaces. Further removal was observed for one of the samples for a second wipe.
\\

\noindent
\underline{Time dependence:}\\
Figure~\ref{fig:time_dependence} shows the
time dependence of the daily counting rates, registered during 112 days of measurements with copper sample 1. It was observed from 2024/05/28 to 2024/12/04 (176 days) or 0.88 $\rm ^{210}Po$ mean live times. The average rates and their statistical errors are given in the legend of figure~\ref{fig:time_dependence}. 
The average rates, observed after the first and second cleaning, are equal within their statistical uncertainties. For the time analysis, the two data sets were combined.
No time dependence is observed for the post-cleaning rates. 
The background-subtracted net rate, observed for copper sample 1, corresponds to a $\rm ^{210}Po$ post-cleaning surface activity of 360 $\mu$Bq.
The simulation error for the rate-to-activity conversion has not been determined because the activity value does not enter into the time analysis.

\begin{figure}[bt]
\includegraphics[width=13cm]{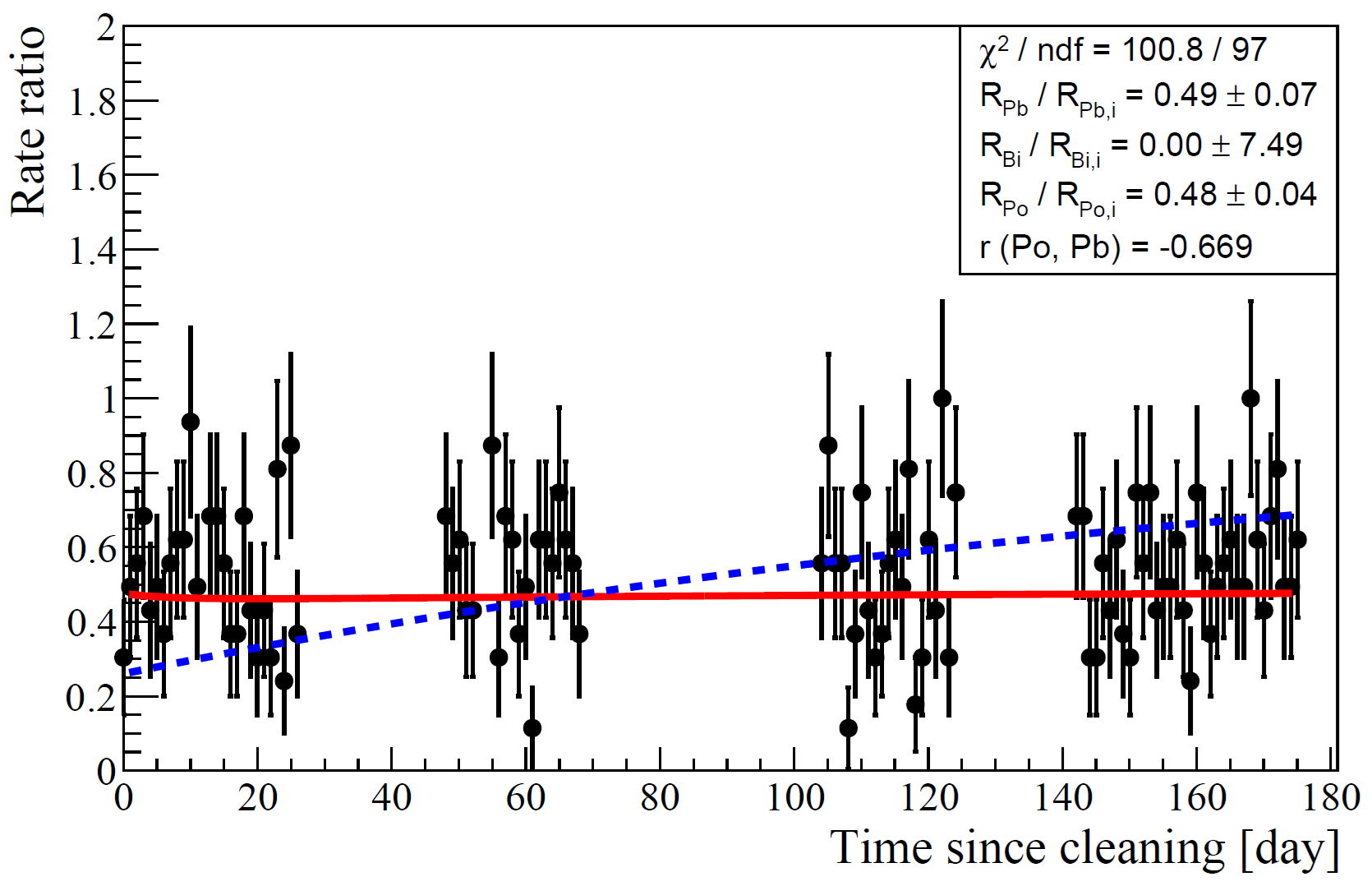}
\caption{\label{fig:time_fit} Time dependence of the background subtracted $\rm ^{210}Po$ cleaned over uncleaned  counting rate ratio for copper sample 1.
The point-wise errors account for the subtraction. The red line and numbers in the fit box show the resulting activity ratios for floating $\rm ^{210}Pb$, $\rm ^{210}Bi$ and $\rm ^{210}Po$ rate ratios. The blue fit line was obtained fixing the  $\rm ^{210}Pb$ and $\rm ^{210}Bi$ activity ratios to 1 (assume no cleaning effect) and floating only the $\rm ^{210}Po$ ratio.
}
\end{figure}

Figure~\ref{fig:time_fit} shows the time dependence of the background subtracted $\rm ^{210}Po$ cleaned over uncleaned counting rate ratio. To understand what quantitative constraints can be put on the removal of $\rm ^{210}Pb$, the data was fitted with equation~\ref{eq:po_decay}.
The red line in figure~\ref{fig:time_fit} was obtained with 
the $\rm A_{Pb}(0)/ A_{Pb, i}$ and $\rm A_{Po}(0)/ A_{Po, i}$ rate ratios free-floating, while $\rm A_{Bi}(0)/ A_{Bi, i}$ 
was constrained to be larger than zero.
The fit results, their uncertainty, and the Po-Pb correlation coefficient are shown in the inset of figure~\ref{fig:time_fit}.
We interpret these ratios as measures of the respective removal fractions. The fit returns equal cleaning efficiencies of $\rm 0.49\pm 0.07$ and $\rm 0.48\pm 0.04$ for $\rm ^{210}Pb$ and $\rm ^{210}Po$, respectively. Removal of $\rm ^{210}Bi$ cannot be determined in this way, as can be seen from the large error of that ratio. 
The fit quality is good. 
Taking into account the correlation of the fit parameters, the ratio of lead to polonium cleaning efficiencies is found to be $\rm 1.02\pm 0.20$, consistent with one.
We conclude from this observation that about half of the surface $\rm ^{210}Pb$ is loosely attached to the surface and the rest implanted or bound. 
Removal of the fit constraint on $\rm A_{Bi}(0)/ A_{Bi, i}$ leads to very similar results for $\rm ^{210}Pb$ and $\rm^{210}Po$ cleaning efficiencies: $\rm 0.50\pm 0.08$ and $\rm 0.49\pm 0.10$, respectively. The fit quality is also good: $\rm \chi^2/ndf=100.7/97$.
We further tested the alternative hypothesis that all $\rm ^{210}Pb$ is left on the surface of copper sample 1 after the two cleanings. The fit was repeated with only
$\rm A_{Po}(0)/ A_{Po, i}$
free floating and $\rm \frac{A_{Pb}(0)}{A_{Pb, i}}=\frac{A_{Bi}(0)}{A_{Bi, i}}=1$ fixed. The blue dashed line in figure~\ref{fig:time_fit} shows the resulting fit. It has $\rm \chi^2/ndf=160.4\; /\; 99$, or 60 units worse than the fit with free floating removal fractions. The hypothesis that all $\rm ^{210}Pb$ and $\rm ^{210}Bi$ is left behind by acetone wiping is disfavored at the $\rm 7.7\cdot \sigma$ level.


\section{Conclusion}
\label{sec:conclusion}

The removal of radon decay products $\rm ^{210}Pb$ and $\rm ^{210}Po$ from copper and silicon surfaces by means of acetone wiping was tested. We find that about half of the surface contamination can be removed by this simple measure. We further find that, within the experimental error, $\rm ^{210}Pb$ and $\rm ^{210}Po$ are removed with equal efficiency.
The $\rm ^{210}Bi$ removal efficiency could not be determined, however, for most practical low background applications its value is irrelevant.

\bmhead{Acknowledgements}
The authors thank their nEXO and XLZD collaborators for their interest and useful discussions. The authors thank J. Busenitz for useful discussions and critical reading of the manuscript.

\bmhead{Statements and Declarations}
This work was supported by the US Department of Energy under grants DE-FG02-01ER41166 and DE-SC0012447. The data sets obtained during this study are available from the corresponding author on a reasonable request.


\bibliography{biblio.bib}   

\end{document}